\documentclass[pra,aps,showpacs,priprent,twocolumn]{revtex4-1}
\usepackage{amsmath}
\usepackage{amssymb}
\usepackage{graphicx}
\usepackage{epstopdf}
\usepackage{graphicx}
\usepackage{bm}
\usepackage{mathtools}
\newcommand{\ket}[1]{|#1\rangle}
\newcommand{\bra}[1]{\langle #1|}
\usepackage[bookmarks=false]{hyperref}
\hypersetup{colorlinks=true,citecolor=blue,linkcolor=blue,urlcolor=blue,pdfstartview=FitH,bookmarksopen=true}
\usepackage{xcolor}
\usepackage{soul}
\usepackage{times,txfonts}
\setstcolor{blue}
\setulcolor{red}

\begin{document}

\title{General approach for constructing Hamiltonians for nonadiabatic holonomic quantum computation}
\author{P. Z. Zhao}
\affiliation{Department of Physics, Shandong University, Jinan 250100, China}
\author{K. Z. Li}
\affiliation{Department of Physics, Shandong University, Jinan 250100, China}
\author{G. F. Xu}
\affiliation{Department of Physics, Shandong University, Jinan 250100, China}
\author{D. M. Tong}
\email{tdm@sdu.edu.cn}
\affiliation{Department of Physics, Shandong University, Jinan 250100, China}
\date{\today}

\begin{abstract}
The main challenges in achieving high-fidelity quantum gates are to reduce the influence of control errors caused by imperfect Hamiltonians and the influence of decoherence caused by environment noise. To overcome control errors, a promising proposal is nonadiabatic holonomic quantum computation, which has attracted much attention in both theories and experiments. While the merit of holonomic operations resisting control errors has been well exploited, an important issue following is how to shorten the evolution time needed for realizing a holonomic gate so as to avoid the influence of environment noise as much as possible. In this paper, we put forward a general approach of constructing Hamiltonians for nonadiabatic holonomic quantum computation, which makes it possible to minimize the evolution time and might open a new horizon for the realistic implementation of nonadiabatic holonomic quantum computation.
\end{abstract}
\maketitle

\section{Introduction}

In 1984, Berry found that a quantum system in a nondegenerate eigenstate, evolving adiabatically and cyclically along a circuit in the parameter space, can acquire a geometric phase in addition to a dynamical phase \cite{Berry}. The notion of geometric phases was then extended to quantum systems in degenerate eigenstates \cite{Wilczek}, in nonadiabatic evolution  \cite{Aharonov,Anandan}, and in mixed states \cite{Uhlmann,Sjoqvist2000,Tong}. Geometric phases are only dependent on the geometric aspect of evolution paths but independent of the evolution details so that they are robust against the control errors that change the evolution rates but keep the evolution paths unchanged. Due to this merit, geometric phases are applied to quantum computation, making quantum gates possess intrinsic robustness against control errors.

The early schemes of geometric quantum computation are based on adiabatic Abelian geometric phases \cite{Berry} or adiabatic non-Abelian geometric phases \cite{Wilczek}, and they are usually known as adiabatic geometric quantum computation \cite{Jones} or adiabatic holonomic quantum computation  \cite{Zanardi99,Duan}. Since adiabatic geometric or adiabatic holonomic  quantum computation requires quantum systems to undergo adiabatic evolution, which makes adiabatic geometric gates vulnerable  to environment-induced decoherence, nonadiabatic geometric quantum computation \cite{WangXB,Zhu} based on nonadiabatic Abelian geometric phases \cite{Aharonov} and nonadiabatic holonomic quantum computation \cite{Sjoqvist2012,Xu2012} based on nonadiabatic non-Abelian geometric phases \cite{Anandan} were proposed. Nonadiabatic holonomic quantum computation shares all the holonomic nature of its adiabatic counterpart but avoids the long run-time requirement. Due to the merits of both its robustness against control errors and its rapidity without the speed limit of the adiabatic evolution, nonadiabatic holonomic quantum computation has received
increasing attention. \cite{Xu2015,Sjovist2016,Sjovist2016PRA,Liang,Zhang,Mousolou2014,Xue,You,Zhao2017,Zhao,Mousolou2017,
Zhao2018,Chen2018,Zhao2019,Sjoqvist2019,Long,Long2017,Peng,Abdumalikov,Xu2018,Danilin,Yin,Egger,Duan2014,
Arroyo,Sekiguchi,Zhou,Nagata,Ishida}.

The first protocol of nonadiabatic holonomic quantum computation \cite{Sjoqvist2012,Xu2012} is based on a three-level quantum system driven by two resonant laser pulses. The schemes based on the original protocol need to combine two sequentially implemented gates to realize a general one-qubit gate. To simplify the operations, the single-shot protocol of nonadiabatic holonomic quantum computation \cite{Xu2015,Sjovist2016} and the single-loop protocol of nonadiabatic holonomic quantum computation \cite{Sjovist2016PRA}, were proposed. Based on the improved protocols, one can directly realize an arbitrary holonomic one-qubit gate with a single-shot implementation, avoiding the extra work of combining two gates into one.
Up to now, the original protocol as well as its reforms have triggered many theoretical schemes \cite{Liang,Zhang,Mousolou2014,Xue,You,Zhao2017,Zhao,Mousolou2017,Zhao2018,Chen2018,Zhao2019,Sjoqvist2019} and experimental demonstrations \cite{Long,Long2017,Peng,Abdumalikov,Xu2018,Danilin,Yin,Egger,Duan2014,
Arroyo,Sekiguchi,Zhou,Nagata,Ishida} of nonadiabatic holonomic quantum computation. Some other progress related to this topic can be seen in Refs. \cite{ZhangSR,Song,XiaSR,Pyshkin,Liu,XiaPRA,Wu,RMP,Xia}, too.

While the merit of holonomic operations, i.e., the robustness against control errors, has been exploited in the schemes of quantum computation, an important issue following is how to further shorten the evolution time needed for realizing a holonomic gate such that the harmfulness of environment noise is possibly reduced. This is nontrivial work, since the Hamiltonian for realizing holonomic quantum gates needs to satisfy two basic conditions: the cyclic evolution condition and the parallel transport condition, which strongly restrict the evolution time. For example,
the evolution time $\tau$ in the previous protocols must satisfy $\int^{\tau}_{0}\Omega(t)dt=\pi$, where $\Omega(t)$ is the envelope of laser pulses. It implies  $\tau\sim\pi/\Omega$, where $\Omega$ is the average modulus of laser pulse envelopes.
The challenge in achieving high-fidelity holonomic quantum gates is to design the Hamiltonians that not only can make the evolution of the quantum system satisfy both the cyclic evolution and the parallel transport conditions but also can accomplish the gate operations within a possibly short time. So far there has not been a general approach to design such Hamiltonians that can minimize the evolution time.

In this paper, we put forward a general approach of constructing Hamiltonians for nonadiabatic holonomic quantum computation. By using the approach, one can choose a desired Hamiltonian such that holonomic gates can be accomplished within a short evolution time, and hence make the quantum gates not only maintain robustness against control errors but also reduce the effect of environment noise.

\section{Approach}

To make our statement clear, we first recall the basic idea of nonadiabatic holonomic quantum computation \cite{Sjoqvist2012,Xu2012}. Nonadiabatic holonomic gates are realized by using a quantum system with a subspace satisfying both the cyclic evolution and parallel transport conditions.
Consider an $N-$dimensional quantum system governed by the Hamiltonian $H(t)$. $\ket{\phi_k(t)}$ represent $N$ orthonormal solutions of the Schr\"{o}dinger equation, i.e., $i\ket{\dot{\phi}_k(t)}=H(t)\ket{\phi_k(t)}$, $k=1,2,\cdots,N$. If there exists an $L-$dimensional subspace   $\{\ket{\phi_k(t)}\}_{k=1}^L$  that satisfies the two conditions:
\begin{align}
&\sum_{k=1}^L |\phi_{k}(\tau)\rangle \langle
\phi_{k}(\tau)|=\sum_{k=1}^L |\phi_{k}(0)\rangle \langle \phi_{k}(0)|,\label{T1}\\
&\langle\phi_{k}(t)|\dot{\phi}_{l}(t)\rangle=0,\ k,l=1,...,L,\label{T2}
\end{align}
then the unitary transformation $U(\tau)$ with $\ket{\phi_k(\tau)}=U(\tau)\ket{\phi_k(0)}$ is a holonomic gate on the $L-$dimensional subspace spanned by $\{ \ket{\phi_k(0)} \}_{k=1}^L$. Here, $\tau$ is the evolution period. This gate is only dependent on evolution paths but independent of evolution details, being robust against control errors.

With the aid of this idea, we may now derive the Hamiltonians that can realize nonadiabatic holonomic quantum computation, i.e., the Hamiltonians that satisfy conditions (\ref{T1}) and (\ref{T2}). For our purpose, we only need to consider a $(L+1)$-dimensional quantum system without the loss of generality.
We use $\{\ket{\nu_1(t)},\ket{\nu_2(t)},\cdots, \ket{\nu_{L+1}(t)}\}$ with $\ket{\nu_k(\tau)}= \ket{\nu_k(0)}$  to represent a set of bases in the Hilbert space, which need not be the solutions of the Schr\"{o}dinger equation, but only a set of auxiliary vectors. We then let
\begin{align}
&\ket{\phi_{k}(t)}=\sum^{L}_{i=1}C_{ik}(t)\ket{\nu_{i}(t)},~~k=1,2,\cdots,L,\label{T3}\\
&\ket{\phi_{L+1}(t)}=e^{i\gamma(t)}\ket{\nu_{L+1}(t)},\label{T4}
\end{align}
where the time dependent coefficients $C_{ik}(t)$ are elements of the $L\times L$ matrix $C(t)$,  defined by
\begin{align}
C(t)=\mathbf{T}e^{i\int^{t}_{0}A(t^{\prime})dt^{\prime}},\label{T6}
\end{align}
with  $A_{ij}(t)=i\bra{\nu_{i}(t)}\dot{\nu}_{j}(t)\rangle$, and $\gamma(t)$ is a real function of $t$ with $\gamma(0)=0$. Note that Eq. (\ref{T6}) is not a result of substituting Eq. (\ref{T3}) into the Schr\"{o}dinger equation but is a requirement for  the quantum system to satisfy conditions (\ref{T2}). Clearly, there are  $\ket{\nu_k(\tau)}= \ket{\nu_k(0)}=\ket{\phi_{k}(0)}$ for $k=1,2,\cdots,L+1$, and $\ket{\phi_{k}(\tau)}=\sum^{L}_{i=1}C_{ik}(\tau)\ket{\phi_{i}(0)}$ for $k=1,2,\cdots,L$.

It is obvious that $\{\ket{\phi_1(t)}, \ket{\phi_2(t)},\cdots, \ket{\phi_L(t)}\}$   satisfy the cyclic condition (\ref{T1}). Besides, by directly substituting Eqs. (\ref{T3}) and (\ref{T6}) into Eq. (\ref{T2}), it is easy to verify that they satisfy the parallel transport condition (\ref{T2}) too.

We now construct Hamiltonian $H(t)$ such that $\ket{\phi_1(t)}, \ket{\phi_2(t)}, \cdots, \ket{\phi_{L+1}(t)}$  are the solutions of Schr\"{o}dinger equation, i.e., $i\ket{\dot{\phi}_k(t)}=H(t)\ket{\phi_k(t)}$, $k=1,2,\cdots,N$. For this, we only need to let
\begin{align}
H(t)=i\sum^{L+1}_{k=1}\ket{\dot{\phi}_k(t)}\bra{\phi_k(t)}.\label{T5}
\end{align}
Substituting Eqs. (\ref{T3}) and (\ref{T4}) into (\ref{T5}), we have
\begin{align}\label{eqTong}
H(t)=&i\sum^{L}_{i,j,k=1} \left(C_{ik}(t)\ket{\nu_{i}(t)}\right)^\prime\left( C_{jk}(t)\ket{\nu_{j}(t)}\right)^\dag \notag\\
&+\left(e^{i\gamma(t)}\ket{\nu_{L+1}(t)}\right)^\prime\left(e^{i\gamma(t)}\ket{\nu_{L+1}(t)}\right)^\dag \notag\\
=&i\sum^{L}_{i,j} \left(\dot{C}(t)C^\dag(t)\right)_{ij}\ket{\nu_{i}(t)}\bra{\nu_{j}(t)}\notag\\
&+i\sum^{L}_{i,j} \left(C(t)C^\dag(t)\right)_{ij}\ket{\dot{\nu}_{i}(t)}\bra{\nu_{j}(t)}\notag\\
&+i\dot{\gamma}(t)\ket{\nu_{L+1}(t)}\bra{\nu_{L+1}(t)}+\ket{\dot{\nu}_{L+1}(t)}\bra{\nu_{L+1}(t)}.
\end{align}
By using the relations $C(t)C^\dag(t)=I$, $\dot{C}(t)C^\dag(t)=iA(t)$,  and $A_{ij}(t)= i\bra{\nu_{i}(t)}\dot{\nu}_{j}(t)\rangle $, we may finally obtain
\begin{align}\label{eq2}
H(t)=&\left[i\sum^{L}_{i=1}\bra{\nu_{i}(t)}\dot{\nu}_{L+1}(t)\rangle\ket{\nu_{i}(t)}\bra{\nu_{L+1}(t)}+\mathrm{H.c.}\right]\notag\\
&+\big[i\bra{\nu_{L+1}(t)}\dot{\nu}_{L+1}(t)\rangle-\dot{\gamma}(t)\big]\ket{\nu_{L+1}(t)}\bra{\nu_{L+1}(t)},
\end{align}
where $\mathrm{H.c.}$ represents the Hermitian conjugate terms.

The above calculations show that starting from an arbitrary set of auxiliary bases  $\{\ket{\nu_{1}(t)}, \ket{\nu_{2}(t)},\cdots, \ket{\nu_{L+1}(t)}    \}$ with $\ket{\nu_k(\tau)}= \ket{\nu_k(0)}$, one may define a Hamiltonian by using formula (\ref{eq2}), of which the Schr\"{o}dinger equation must be satisfied by  $\ket{\phi_1(t)}, \ket{\phi_2(t)}\cdots,\ket{\phi_{L+1}(t)}$, defined by Eqs. (\ref{T3}) and (\ref{T4}).  The subspace, $\mathcal{S}_L(t)= \text{Span}\{\ket{\phi_1(t),\phi_2(t),\cdots, \phi_L(t)}\}$, naturally satisfies both the cyclic evolution and parallel transport conditions, and therefore $\mathcal{S}_L(0)=\text{Span}\{\ket{\phi_1(0),\phi_2(0),\cdots, \phi_L(0)}\}$, i.e.,  $\text{Span}\{\ket{\nu_1(0),\nu_2(0),\cdots, \nu_L(0)}\}$, can be taken as the computational space of  nonadiabatic holonomic computation.  In this case, the unitary operator acting on the subspace $\mathcal{S}_{L}(0)$ is simply given by
\begin{align}\label{eq3}
U(\tau)=C(\tau)=\mathbf{T}e^{i\int^{\tau}_{0}A(t)dt}.
\end{align}

So far, we have put forward a general approach of constructing Hamiltonians, expressed as formula (\ref{eq2}), for nonadiabatic holonomic quantum computation.

Before going to applications, we would like to add the following remark. To realize an $L\times L$ nonadiabatic holonomic gate, we need an $N\geq(L+1)-$dimensional space with the remanent $(N-L)-$dimensional subspace acting as an auxiliary. The simplest choice is $N=L+1$, as did in the above. Certainly, one can also take a larger $N$, i.e., a higher-dimensional auxiliary subspace, but a higher-dimensional auxiliary subspace will make the Hamiltonian include more extra couplings between the $L$-dimensional computational subspace and the $(N-L)$-dimensional auxiliary subspace. This is complicated in the practical implementation. Therefore, we choose an $N=(L+1)-$dimensional space. However, the approach can be easily extended to higher-dimensional cases.

\section{Applications}

The above discussion provides an effective approach to realizing nonadiabatic holonomic quantum computation, which makes it possible to minimize the evolution time  needed for realizing quantum gates.
To show its usefulness, as an example, we will give a universal set of nonadiabatic holonomic gates, i.e., arbitrary one-qubit gates and a non-trivial two-qubit gate, from which one will see that the Hamiltonians used in the previous schemes are only special cases of the formula in Eq. (\ref{eq2}) and an alternative choice of Hamiltonian can markedly reduce the evolution time.

\subsection{Application to one-qubit gate}

To realize an arbitrary one-qubit gate, we consider a three-level system consisting of qubit states $\{\ket{0},\ket{1}\}$ and an auxiliary state $\ket{e}$.  One choice of the auxiliary bases \cite{Tong1} can be
\begin{align}\label{T7}
\ket{\nu_{1}(t)}=&\cos\frac{\theta}{2}\ket{0}+\sin\frac{\theta}{2}e^{i\varphi}\ket{1},
\notag\\
\ket{\nu_{2}(t)}=&\cos\frac{\alpha(t)}{2}\sin\frac{\theta}{2}e^{-i\varphi}\ket{0}
\notag\\
&-\cos\frac{\alpha(t)}{2}\cos\frac{\theta}{2}\ket{1}+\sin\frac{\alpha(t)}{2}e^{i\beta(t)}\ket{e},
\notag\\
\ket{\nu_{3}(t)}=&\sin\frac{\alpha(t)}{2}\sin\frac{\theta}{2}e^{-i[\varphi+\beta(t)]}\ket{0}
\notag\\
&-\sin\frac{\alpha(t)}{2}\cos\frac{\theta}{2}e^{-i\beta(t)}\ket{1}-\cos\frac{\alpha(t)}{2}\ket{e},
\end{align}
where $\theta$ and $\varphi$ are time-independent parameters, and $\alpha(t)$ and $\beta(t)$ are time-dependent parameters with $\alpha(0)=\alpha(\tau)=0$. Clearly, $\mathcal{S}_{L}(t)=\mathrm{Span}\{\ket{\nu_{1}(t)},\ket{\nu_{2}(t)}\}$ undergoes  a cyclic evolution with period time $\tau$, and hence the initial subspace $\mathcal{S}_{L}(0)=\mathrm{Span}\{\ket{\nu_{1}(0)},\ket{\nu_{2}(0)}\}
=\mathrm{Span}\{\ket{0},\ket{1}\}$ can be taken as the computational space.

\begin{figure}[t]
\begin{center}
\includegraphics[scale=0.25]{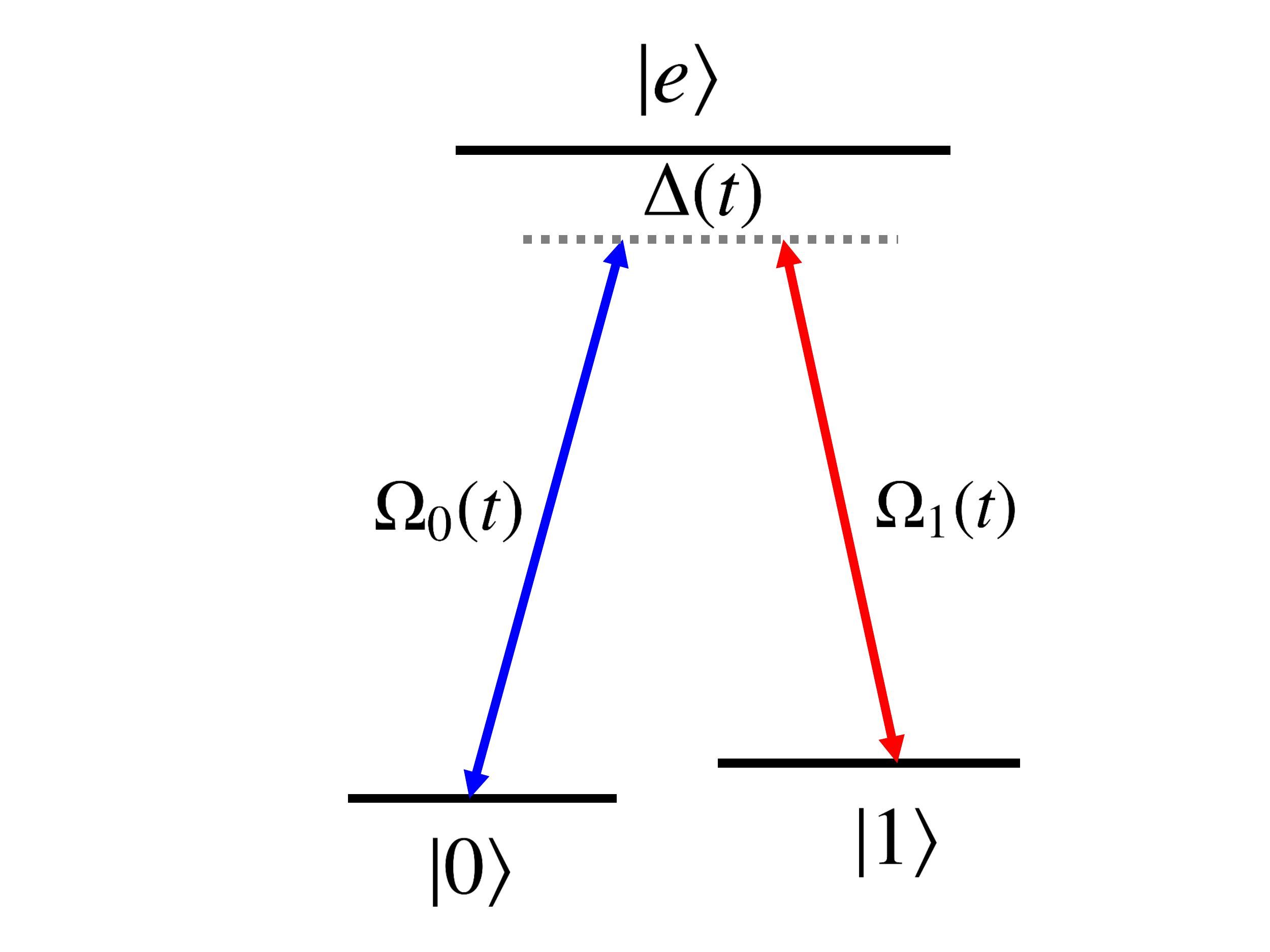}
\end{center}
\caption{(Color online) Level configuration of the three-level system driven by two lasers. \label{Figure1}}
\end{figure}

We take $\gamma(t)$ to satisfy $\dot{\gamma}(t)=\dot{\beta}(t)[3+\cos\alpha(t)]/2$ so as to eliminate the undesirable coupling \cite{Tong2}. In this case, the Hamiltonian given by Eq. (\ref{eq2}) reads
\begin{align}\label{eq4}
H(t)=&-\dot{\beta}(t)[1+\cos\alpha(t)]\ket{e}\bra{e}
\notag\\
&+\frac{1}{2}\Bigg\{\left[i\dot{\alpha}(t)+\dot{\beta}(t)\sin\alpha(t)\right]\sin\frac{\theta}{2} e^{i[\varphi+\beta(t)]}\ket{e}\bra{0}
\notag\\
&-\left[i\dot{\alpha}(t)+\dot{\beta}(t)\sin\alpha(t)\right]\cos\frac{\theta}{2} e^{i\beta(t)}\ket{e}\bra{1}+\mathrm{H.c.}\Bigg\}.
\end{align}
It means that the three-level system is driven by two lasers with Rabi frequencies $\Omega_0(t)=[i\dot{\alpha}(t)+\dot{\beta}(t)\sin\alpha(t)]\sin(\theta/2)\exp\{i[\varphi+\beta(t)]\}/2$ and $\Omega_1(t)=-[i\dot{\alpha}(t)+\dot{\beta}(t)\sin\alpha(t)]\cos(\theta/2)\exp[i\beta(t)]/2$, respectively, and with the same detuning $\Delta(t)=-\dot{\beta}(t)[1+\cos\alpha(t)]$. $H(t)$ can be briefly written as
\begin{align}\label{eq4T}
H(t)=\Delta(t)\ket{e}\bra{e}+[\Omega_0(t)\ket{e}\bra{0}+\Omega_1(t)\ket{e}\bra{1}+\mathrm{H.c.}],
\end{align}
of which the level configuration is shown as Fig. \ref{Figure1}.

The quantum system governed by $H(t)$ undergoes a unitary evolution, and at the time $t=\tau$, the unitary operator acting on the computational space can be obtained by using Eq. (\ref{eq3}). It reads
\begin{align}\label{eq5}
U(\tau)=e^{i\phi(\tau)\boldsymbol{\mathrm{n}\cdot\sigma}/2},
\end{align}
where $\boldsymbol{\mathrm{n}}=(\sin\theta\cos\varphi,\sin\theta\sin\varphi,\cos\theta)$ is a unit vector,
$\boldsymbol{\sigma}=(\sigma_{x},\sigma_{y},\sigma_{z})$ are Pauli operators, and
\begin{align}\label{T9}
 \phi(\tau)=\frac{1}{2}\int^{\tau}_{0}[1-\cos\alpha(t)]\dot{\beta}(t)dt
 \end{align}
is a rotation angle with $0\leq \phi(\tau)<2\pi$. Here, a trivial global phase in $U(\tau)$ has been removed from Eq. (\ref{eq5}). $U(\tau)$ plays an arbitrary one-qubit nonadiabatic holonomic gate, as both the rotation axis $\boldsymbol{\mathrm{n}}$ and rotation angle $\phi(\tau)$ can be arbitrarily taken.

\begin{figure}[t]
\begin{center}
\includegraphics[scale=0.25]{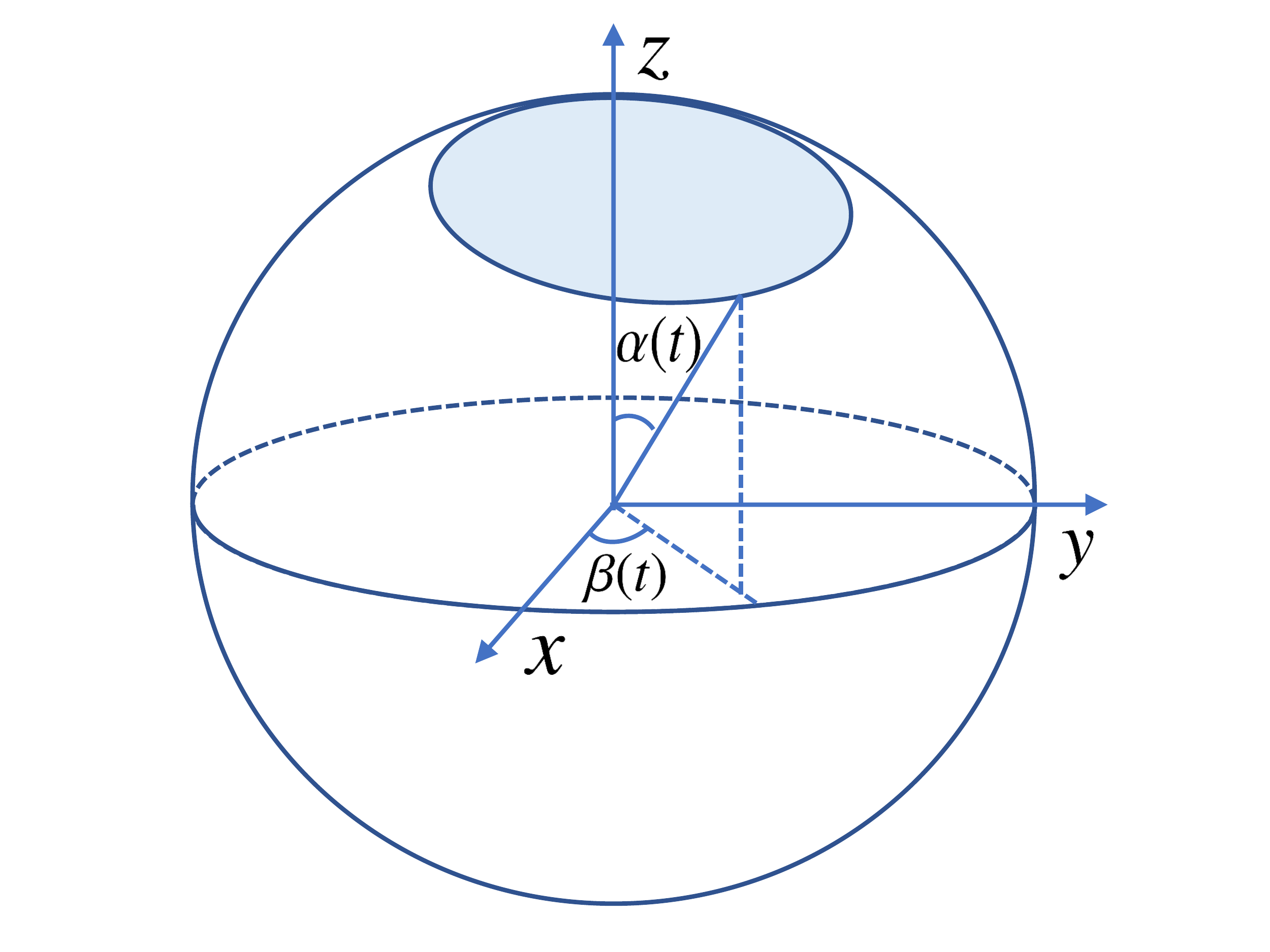}
\end{center}
\caption{(Color online) The unit sphere describing the changes of  $\alpha(t)$ and  $\beta(t)$ with time $t$.   \label{Figure2}}
\end{figure}

If we take the parameters  $\alpha(t)$ and  $\beta(t)$ as the polar angle and azimuthal angle of a spherical coordinate system, respectively, then  $[\alpha(t),\beta(t)]$ represents a point in a unit 2-sphere, and it traces a closed path $C$ in the unit sphere when the time changes from $t=0$ to $t=\tau$ (see Fig. \ref{Figure2}).
It is interesting to note that $\phi(\tau)$ can be recast as
\begin{align}\label{T12}
 \phi(\tau)=\frac{1}{2}\oint_C(1-\cos\alpha)d\beta,
\end{align}
which is just equal to the half of the solid angle enclosed by the path $C$ in the space of parameters.
Equation (\ref{T12}) shows that the phase  $\phi(\tau)$ is only dependent on the path traced by  $\alpha(t)$ and  $\beta(t)$ but independent of the changing rate of them, being robustness against control errors.

The parameters $\alpha(t)$ and $\beta(t)$ are two undetermined functions of $t$. Different choices of them  correspond to different paths $C$,  which result in different Hamiltonians, i.e. different protocols of holonomic quantum computation.
To realize a given gate, there are infinitely many protocols of choosing the functions $\alpha(t)$ and $\beta(t)$, but the evolution time needed by using different protocols may be quite different. This provides a way to find  optimal evolution paths. To illustrate this point, we take the $\pi/8$-gate, i.e., $U(\tau)=\exp[i\phi(\tau)\boldsymbol{\mathrm{n}\cdot\sigma}/2]$ with $\phi(\tau)=\pi/8$, as an example.
First, we can take $[\alpha(t),\beta(t)]$ starting from the north pole along the great circle with $\beta(t)=0$ to the south pole, and then return back to the north pole from the south pole along another great circle with $\beta(t)=\pi/8$. In this case, the piecewise Hamiltonian is given by Eq. (\ref{eq4}) as
$H(t)=\dot{\alpha}(t)\{\sin(\theta/2)\exp[i(\varphi+\pi/2)]\ket{e}\bra{0}-\cos(\theta/2) \exp(i\pi/2) \ket{e}\bra{1}\}/2+\mathrm{H.c.}$
for $0\leq t\leq\tau/2$, and $H(t)=\dot{\alpha}(t)\{\sin(\theta/2) \exp[i(\varphi+5\pi/8)]\ket{e}\bra{0}-\cos(\theta/2)\exp[i(5\pi/8)]\ket{e}\bra{1}\}/2+\mathrm{H.c.}$ for  $\tau/2<t\leq\tau$. Here, $\dot{\alpha}(t)/2$ plays  the role of laser pulse envelope $\Omega(t)$ and can be expressed as $\alpha(t)=2\int^{t}_{0}\Omega(t^{\prime})dt^{\prime}$.
This is just the one-loop protocol \cite{Sjovist2016PRA}. In this case, the length of the path, which is enclosed by two meridians, is $l_{C_1}=2\pi$. Note that the path used in the original protocol  \cite{Sjoqvist2012,Xu2012} is also corresponding to the one enclosed by two meridians and has the same length.

Alternatively, we can take $[\alpha(t),~\beta(t)]$ starting from the north pole, along the first great circle with $\beta(t)=0$, to the point $(\pi/2,~0)$, then along the second great circle with $\alpha(t)=\pi/2$ to the point $(\pi/2,~\pi/4)$, and finally return back to the north pole from the point $(\pi/2,~\pi/4)$ along the third great circle with $\beta(t)=\pi/4$. In this case, the length of the path is $l_{C_2}=5\pi/4$. If the change rates of $\alpha(t)$ and $\beta(t)$, which are determined by the envelope of laser pulses, are same in the two protocols, then there is $\tau_{C_2}=(5/8)\tau_{C_1}$, where $\tau_{C_1}$ and $\tau_{C_2}$ respectively represents the time needed by the previous protocols and by the alternative protocol.

Our approach can help to choose any desired evolution path and minimize the evolution time needed for realizing a holonomic gate. More interestingly, we can estimate the minimum time needed for realizing a general $\phi$-gate,  $U(\tau)=\exp(i\phi\boldsymbol{\mathrm{n}\cdot\sigma}/2)$, by using Eq. (\ref{T12}). Since the state of the quantum system is initially in the computational space spanned by $\{\ket{0},\ket{1}\}$, we have $\alpha(0)=0$ from Eq. (\ref{T7}), which means that  $[\alpha(t),\beta(t)]$ starts from the north pole of the unit sphere. On the other hand, the circumference of a circle is shorter than any other shapes for a given area. Therefore, the shortest path that encloses the solid angle $2\phi$  can be any circle passing through the north pole. One of these circles, for instance, can be expressed as $(\pi-\phi)(1-\cos\alpha(t))-\sqrt{2\pi\phi-\phi^2}\sin\alpha(t)\cos\beta(t)=0$. The length of these circles is $l_{\mathrm{min}}=2\sqrt{2\pi\phi-\phi^2}$. Noting that the length of those paths in the previous protocols \cite{Sjoqvist2012,Xu2012,Sjovist2016PRA} is $2\pi$, the minimum evolution time given by our formalism  can be expressed as $\tau_{\mathrm{min}}/\tau_{C_1}=\sqrt{2\phi/\pi-(\phi/\pi)^2}$. There is always $\tau_{\mathrm{min}}<\tau_{C_1}$ for $0< \phi<\pi$ and $\pi<\phi<2\pi$.

\subsection{Application to two-qubit gate}
To realize a nontrivial two-qubit gate, we choose the auxiliary bases to be
\begin{align}\label{T8}
\ket{\nu_{1}(t)}=&\ket{00},~~\ket{\nu_{2}(t)}=\ket{01},
\notag\\
\ket{\nu_{3}(t)}=&\cos\frac{\theta}{2}\ket{10}+\sin\frac{\theta}{2}e^{i\varphi}\ket{11},
\notag\\
\ket{\nu_{4}(t)}=&\cos\frac{\alpha(t)}{2}\sin\frac{\theta}{2}e^{-i\varphi}\ket{10}
\notag\\
&-\cos\frac{\alpha(t)}{2}\cos\frac{\theta}{2}\ket{11}+\sin\frac{\alpha(t)}{2}e^{i\beta(t)}\ket{ee},
\notag\\
\ket{\nu_{5}(t)}=&\sin\frac{\alpha(t)}{2}\sin\frac{\theta}{2}e^{-i[\varphi+\beta(t)]}\ket{10}
\notag\\
&-\sin\frac{\alpha(t)}{2}\cos\frac{\theta}{2}e^{-i\beta(t)}\ket{11}-\cos\frac{\alpha(t)}{2}\ket{ee},
\end{align}
where $\theta$ and $\varphi$ are the time-independent parameters, and $\alpha(t)$ and $\beta(t)$ are time-dependent parameters with $\alpha(0)=\alpha(\tau)=0$. Clearly, the subspace $\mathcal{S}_{L}(t)=\mathrm{Span}\{\ket{\nu_{1}(t)},\ket{\nu_{2}(t)},\ket{\nu_{3}(t)},\ket{\nu_{4}(t)}\}$ undergoes a cyclic evolution with the period $\tau$, and $\mathcal{S}_{L}(0)$ can be taken as the computational space.
 We further take $\dot{\gamma}(t)=\dot{\beta}(t)[3+\cos\alpha(t)]/2$. From formula (\ref{eq2}), we have the Hamiltonian of the quantum system as
\begin{align}
H(t)=&\frac{1}{2}\Bigg\{\left[i\dot{\alpha}(t)+\dot{\beta}(t)\sin\alpha(t)\right]\sin\frac{\theta}{2} e^{i[\varphi+\beta(t)]}\ket{ee}\bra{10}
\notag\\
&-\left[i\dot{\alpha}(t)+\dot{\beta}(t)\sin\alpha(t)\right]\cos\frac{\theta}{2} e^{i\beta(t)}\ket{ee}\bra{11}+\mathrm{H.c.}\Bigg\}
\notag\\
&-\dot{\beta}(t)[1+\cos\alpha(t)]\ket{ee}\bra{ee}.
\end{align}
In this case, after a cyclic evolution, the unitary operator acting on the computational space  reads
\begin{align}\label{eq9}
U(\tau)=&\ket{0}\bra{0}\otimes I
+e^{-i\phi(\tau)/2}\ket{1}\bra{1}\otimes e^{i\phi(\tau)\boldsymbol{\mathrm{n}\cdot\sigma}/2},
\end{align}
with $\phi(\tau)=\int^{\tau}_{0}[1-\cos\alpha(t)]\dot{\beta}(t)dt/2$.
This is a nontrivial two-qubit nonadiabatic holonomic gate. From Eqs. (\ref{T7}) and (\ref{T8}), we can see that the bases $\{\ket{\nu_{3}(t)},\ket{\nu_{4}(t)},\ket{\nu_{5}(t)}\}$ in Eq. (\ref{T8}) have the same form as  the bases $\{\ket{\nu_{1}(t)},\ket{\nu_{2}(t)},\ket{\nu_{3}(t)}\}$ in Eq. (\ref{T7}) while the bases $\{\ket{\nu_{1}(t)},\ket{\nu_{2}(t)}\}$ in Eq. (\ref{T8}) are invariant. Therefore, we can do further discussions for the two-qubit gate, as done with the one-qubit gates, and can demonstrate all the points mentioned in the one-qubit case.

\section{Conclusion}

We have put forward a general approach of constructing Hamiltonians for nonadiabatic holonomic quantum computation. Starting from an arbitrary set of auxiliary bases $\{\ket{\nu_{1}(t)}, \ket{\nu_{2}(t)},\cdots, \ket{\nu_{L+1}(t)}\}$ with $\ket{\nu_k(\tau)}= \ket{\nu_k(0)}$ and using the formula in Eq. (\ref{eq2}), one may easily write out the Hamiltonian with $\ket{\phi_1(t)}, \ket{\phi_2(t)}\cdots,\ket{\phi_{L+1}(t)}$, defined by Eqs. (\ref{T3}) and (\ref{T4}), being the solutions of Schr\"{o}dinger equation.  The subspace, $\mathcal{S}_L(t)= \text{Span}\{\ket{\phi_1(t),\phi_2(t),\cdots, \phi_L(t)}\}$, naturally satisfies both the cyclic evolution and parallel transport conditions, and therefore $\mathcal{S}_L(0)$  can be taken as the computational space of  nonadiabatic holonomic computation. Our finding greatly simplifies the process of designing the Hamiltonians for holonomic quantum computation, and makes it possible to minimize the evolution time needed for realizing a holonomic gate. To show its application, we have given a universal set of holonomic quantum gates with a much shorter evolution time than the previous protocol, and also demonstrate the shortest paths.

\begin{acknowledgments}
P.Z.Z. acknowledges support from the National Natural Science Foundation of China through Grant No. 11947221. K.Z.L. acknowledges support from the National Natural Science Foundation of China through Grant No. 11575101. D.M.T. acknowledges support from the National Natural Science Foundation of China through Grant No. 11775129.
\end{acknowledgments}


\begin{thebibliography}{99}
\bibitem{Tong1} To serve our purpose, we here chose the auxiliary bases to be those in Eq. (\ref{T7}). However, the auxiliary bases can be arbitrarily chosen according to one's necessity. Each choice corresponds to a different Hamiltonian. For example, they can also be taken as $\ket{\nu_{1}(t)}=\cos\mu(t)\ket{0}-\sin\mu(t)\exp(i\varphi)\ket{1}$, $\ket{\nu_{2}(t)}=\cos\xi(t)\sin\mu(t)\exp(-i\varphi)\ket{0}+\cos\xi(t)\cos\mu(t)\ket{1}-\sin\xi(t)\ket{e}$, $\ket{\nu_{3}(t)}=\sin\xi(t)\sin\mu(t)\exp(-i\varphi)\ket{0}+\sin\xi(t)\cos\mu(t)\ket{1}+\cos\xi(t)\ket{e}$,
where $\mu(0)=0$ and $\xi(0)=\xi(\tau)=0$. In this case, by using Eq. (\ref{eq2}) with  $\gamma(t)=0$, there is $H(t)=-i\left[\dot{\mu}(t)\sin\xi(t)\cos\xi(t)\cos\mu(t)+\dot{\xi}(t)\sin\mu(t)\right]\exp(i\varphi)\ket{e}\bra{0}+i[\dot{\mu}(t)\sin\xi(t)\cos\xi(t)\sin\mu(t)-\dot{\xi}(t)\cos\mu(t)]
\ket{e}\bra{1}-i\dot{\mu}(t)\sin^{2}\xi(t)\exp(i\varphi)\ket{1}\bra{0}+\mathrm{H.c.}$. The operator acting on the computational space $\{\ket{0},\ket{1}\}$ reads $U(\tau)=\exp[i\epsilon(\tau)(\sin\omega\sigma_{x}-\cos\omega\sigma_{y})]$ with $\epsilon(\tau)=\oint_C\cos\xi d\mu$.
\bibitem{Tong2} When Eq. (\ref{T7}) is substituted into Eq. (\ref{eq2}), there is a coupling term between $\ket{0}$ and $\ket{1}$ in the Hamiltonian, reading $\frac{1}{2}[1-\cos\alpha(t)]\{\dot{\gamma}(t)-\dot{\beta}(t)[3+\cos\alpha(t)]/2\}
\sin(\theta/2)\cos(\theta/2)\exp(-i\varphi)\ket{0}\bra{1}+\mathrm{H.c.}$, which is electricdipole forbidden for the hyperfine level states. Such an undesirable coupling can be eliminated by taking $\dot{\gamma}(t)=\dot{\beta}(t)[3+\cos\alpha(t)]/2$.
\end{thebibliography}
\end{document}